\documentclass[runningheads]{cl2emult}

\usepackage{makeidx}  
\usepackage{graphicx} 
\usepackage{subeqnar} 
\usepackage{multicol} 
\usepackage{cropmark} 
\usepackage{eso}      
\makeindex            

\begin{document}
\title*{BMW: a ROSAT-HRI Source Catalogue Obtained with a
Wavelet Transform Detection Algorithm}
\toctitle{BMW: a ROSAT-HRI Source Catalogue Obtained with a
Wavelet Transform Detection Algorithm}
\titlerunning{BMW: a ROSAT-HRI Source Catalogue}
\author{Maria~Rosa~Panzera\inst{1}
\and Sergio Campana\inst{1}
\and Stefano Covino\inst{1}
\and Luigi Guzzo\inst{1}
\and Gian~Luca~Israel\inst{2}
\and Davide Lazzati\inst{1}
\and Roberto Mignani\inst{3}
\and Alberto Moretti\inst{1}
\and Gianpiero Tagliaferri\inst{1}}
\authorrunning{Maria~Rosa~Panzera et al.}
\institute{Osservatorio Astronomico di Brera, Via E. Bianchi 46, I-23807 
Merate, Italy
\and Osservatorio Astronomico di Roma, Via Frascati 33,\\ 
I-00040 Monteporzio Catone, Italy
\and European Southern Observatory, Garching bei Munchen, Germany}

\maketitle              

\begin{abstract}
In collaboration with the
Observatories of Palermo and
Rome and the SAX-SDC we are constructing a multi-site interactive
archive system featuring specific analysis tools.
In this context we developed a detection algorithm based on the
Wavelet Transform (WT) and performed a systematic analysis of all
ROSAT-HRI public data ($\sim$ 3100 observations $+$ 1000 to come).
The WT is specifically suited to detect and characterize
extended sources while properly detecting point sources in very
crowded fields. Moreover, the good angular resolution of HRI images
allows the source extension and position to be accurately determined.

This effort has produced
the BMW (Brera Multiscale Wavelet) catalogue,
with more than 19,000 sources detected at the $\sim 4.2 \,\sigma$ level.
For each source detection we have information on the position, X-ray flux 
and extension.
This allows for instance to select complete samples of extended X-ray
sources such as candidate clusters of galaxies 
or SNR's.  Details about the detection algorithm and the
catalogue can be found in Lazzati et al. 1999, Campana et
al. 1999 and Panzera et al. 2001.  Here we shall present an overview of 
first results from several undergoing projects which make use of the BMW 
catalogue.
\end{abstract}

\section{The Catalogue}
\begin{figure}[!t]
\centering
\includegraphics[angle=0, width=1.\textwidth]{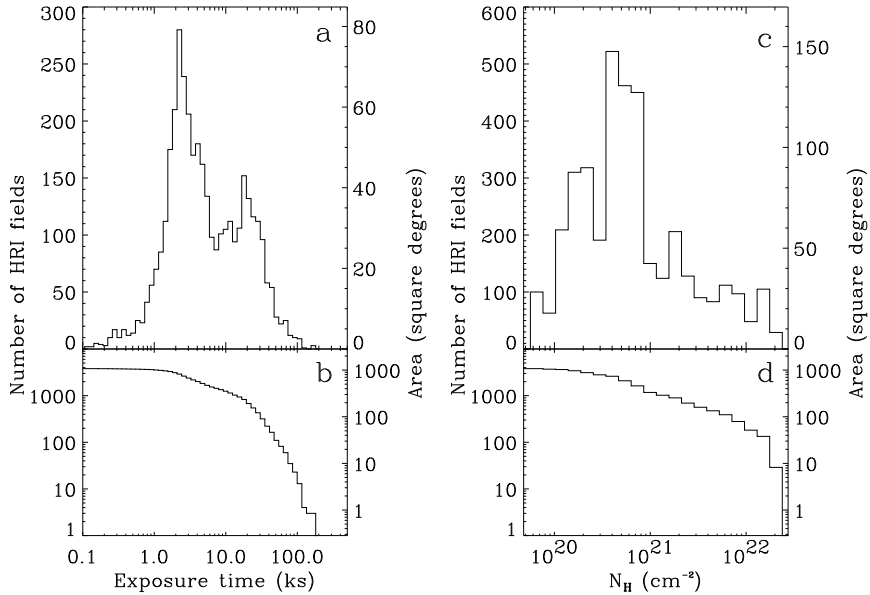}
\caption[]{({\bf a}) Distribution of the HRI exposure times of the HRI
images (all but SNR and calibration fields) that we have analyzed; ({\bf b})
cumulative distribution. ({\bf c}) Distribution of the Galactic column density 
in the same fields; ({\bf d}) cumulative distribution.
}
\label{eps1}
\end{figure}

\begin{figure}[!t]
\centering
\includegraphics[angle=0, width=1.\textwidth]{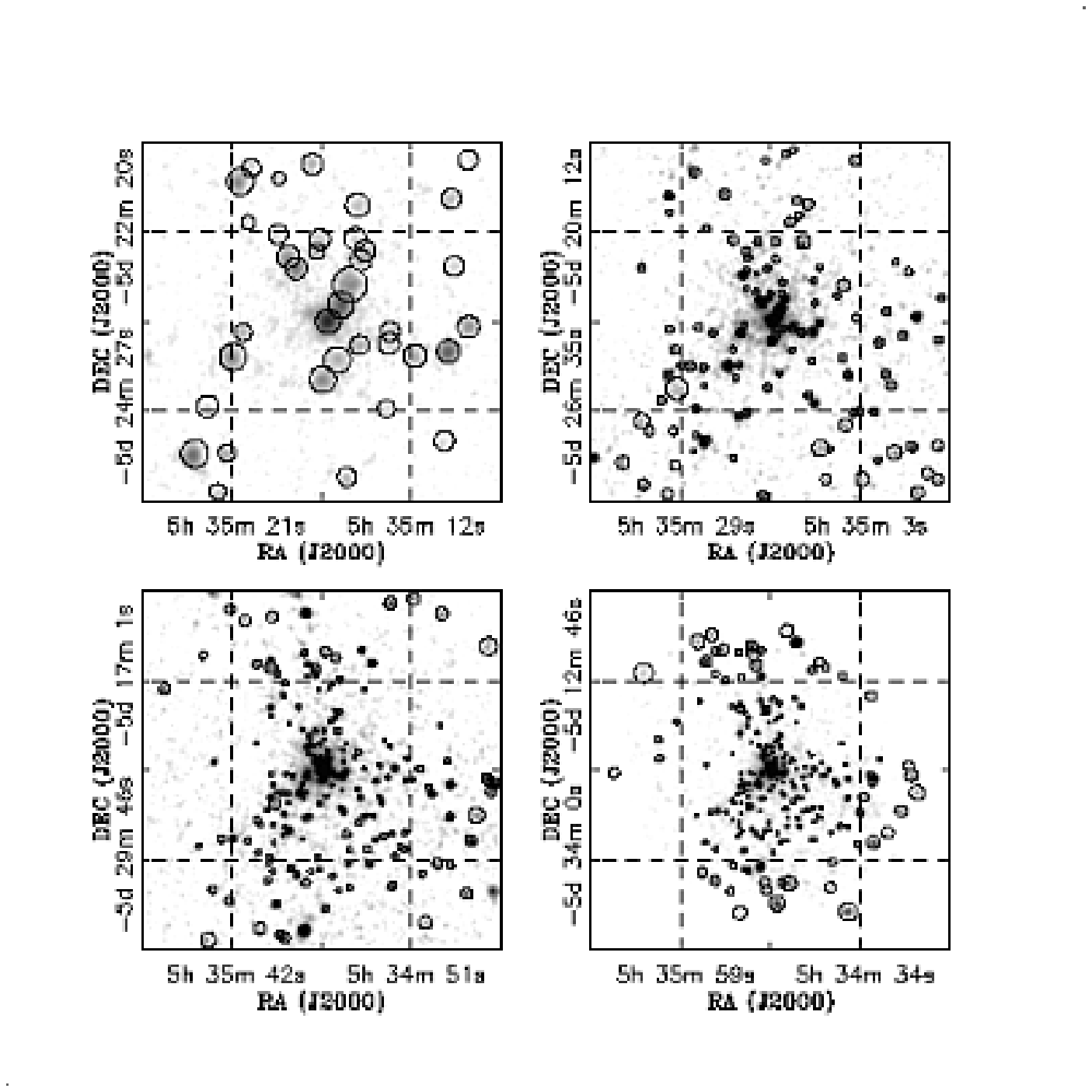}
\caption[]{Source detection in the Trapezium field. The four panels show
the smoothed images at rebin 1, 3, 6 and 10 (i.e. pixel size
of 0.5, 1.5, 3.0 and 5 arcsec, respectively), respectively.
Circles mark X-ray sources; the size of the circles is twice the source 
width ($\sigma$; modeled as a Gaussian).}
\label{eps2}
\end{figure}

The wavelet detection algorithm (WDA) we developed was made suited for a fast
and efficient analysis of images taken with the ROSAT HRI instrument
(see Lazzati et al. 1999 and Campana et al. 1999).
From the automatic analysis of all pointing observations available in the 
ROSAT public archive (at HEASARC and MPE) as of January 1999, we derived the
BMW catalogue with sources detected with a significance 
$\sim\,4.2\,\sigma$ ($\sim$ 19000).
All the detected HRI sources are characterized in flux, size and position.
In Fig.~\ref{eps1} we show the differential and integral distributions
of exposure times and galactic absorptions;
while in Fig.~\ref{eps2} we plot the results of source detection 
applied to the crowded Trapezium field.
At variance with the \lq \lq sliding box'' detection algorithms the WDA
provides also a reliable description of the source extension allowing
for a complete search of e.g. supernova remnants or clusters of galaxies
in the HRI fields.
To assess the source extension we considered all the sources detected in the 
observations that have a star(s) as a target (ROR number beginning with 2):
6013 in 756 HRI fields. 
The distribution of the source width ($\sigma$\/) as a function of
the source off-axis angle has been divided into bins of 1 arcmin 
each in which
we applied a $\sigma$-clipping algorithm on the source width.
This method iteratively discards truly extended sources and provides the mean
value of $\sigma$ for each bin.
We then determined the $3\,\sigma$ dispersion on the mean for each bin.
The mean value plus the $3\,\sigma$ dispersion provide the threshold for
the source extension ({\it dashed line} in Fig.~\ref{eps3}; see also
Rosati et al. 1995).
We conservatively classify a source as extended if it lies more than
$2\,\sigma$ from this limit ({\it filled squares} in Fig.~\ref{eps3}).

More than 1000 new HRI fields have been taken from the ROSAT public
archive and will be analyzed in the next future. 
\section{First Results}
In the following we present the first results obtained using the BMW
catalogue.
\subsection{Search for Periodic Signals}
\begin{figure}[!t]
\centering
\includegraphics[angle=270, width=1.\textwidth]{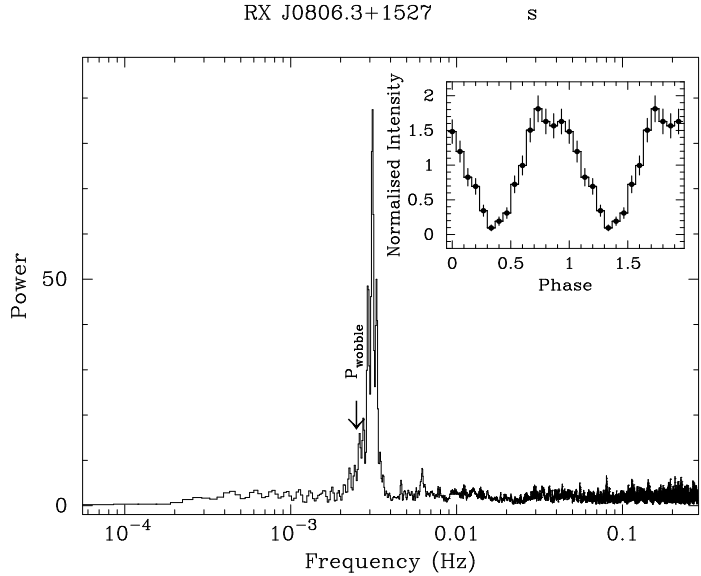}
\caption[]{The average power spectra obtained for RX J0806.3+1527
light curves by using both HRI observations (ROR 300421n00 and ROR 300421a01).
The highly significant peak ($\sim\,9\,\sigma$ based on the fundamental
only) was found at a frequency of 0.0031127 Hz, corresponding to a period of 
321.25 s. The frequency of the wobble is marked with a vertical {\it solid
arrow}. The RX J0806.3+1527 light curves folded at the best period (321.25 s)
is also shown in the little upper right panel}
\label{eps4}
\end{figure}

The BMW catalogue 
contains about 3000 sources with more than 160 photons, which we 
set as the minimum number required to carry out a meaningful search for
periodic signals (see Israel et al. 1998).
In collaboration with the Osservatorio Astronomico di Roma, these
light curves were analyzed in a systematic way by using the
algorithm of Israel \& Stella (1996) for the detection of coherent or 
quasi-coherent signals in the power spectra even in presence of additional
non-Poissonian noise components.
The technique was modified to correct for the spurious effects which 
characterise the ROSAT light curves.
During this systematic search we discovered $\sim$ 321 s pulsations in the
X-ray flux of 1BMW J080622.8 +152732 = RX J0806.3+1527 (Fig.~\ref{eps4}).
Two different HRI observations were obtained with the source
at flux level of 3 and 5\,$\times$\,10$^{-12}$ erg cm$^{-2}$ s$^{-1}$,
respectively. Only a faint $B$\,$=$\,20.5 object is present
within the error
circle, while no optical counterpart is present in the $R$ plate down to a
limiting magnitude of $\sim$\,20. This indicates that the object is
intrinsically blue. The X-ray and optical findings imply that the source
is a relatively distant ($\sim$\,500\,pc)
isolated white dwarf or, a nearby ($\sim$\,10\,pc) isolated
neutron star accreting from the interstellar medium (Israel et al. 1999;
Israel et al. 2001).
\begin{figure}[!t]
\centering
\includegraphics[angle=0, width=1.\textwidth]{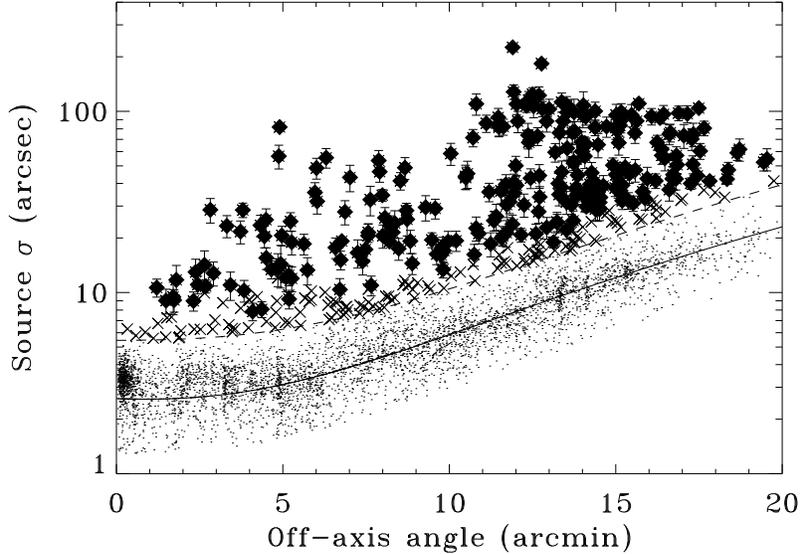}
\caption[]{Source extension ($\sigma$\/) vs. off-axis angle for 6013
sources detected in the HRI fields pointed on stellar targets.
{\it Dashed line}, marks the 3 $\sigma$ extension limit for point-like 
sources; {\it solid line}, computed PSF.
}
\label{eps3}
\end{figure}
\subsection{Search for Clusters of Galaxies}
The evolution of the abundance of massive clusters of galaxies 
represents a key test for the models of large-scale structure formation.   
X-ray observations are the most fruitful (and physically sound) way to 
find such objects at high redshift
($z\,>\,0.5$), as demonstrated in recent years by surveys based 
on the ROSAT PSPC archival data
(e.g. Rosati et al. 1998, Vikhlinin et al. 1998a).  
On the contrary, the ROSAT HRI data archive received poor attention
as a source for cluster searches, due to its lower sensitivity and higher
background. Our first results on BMW cluster candidates are showing
that these problems can be overcome with a clever analysis of the data
(see Campana et al. 1999). 

\begin{figure}[!t]
\begin{tabular}{c}
~~~~~~~~~~~~~~~~{\includegraphics[angle=0, width=0.7\textwidth]{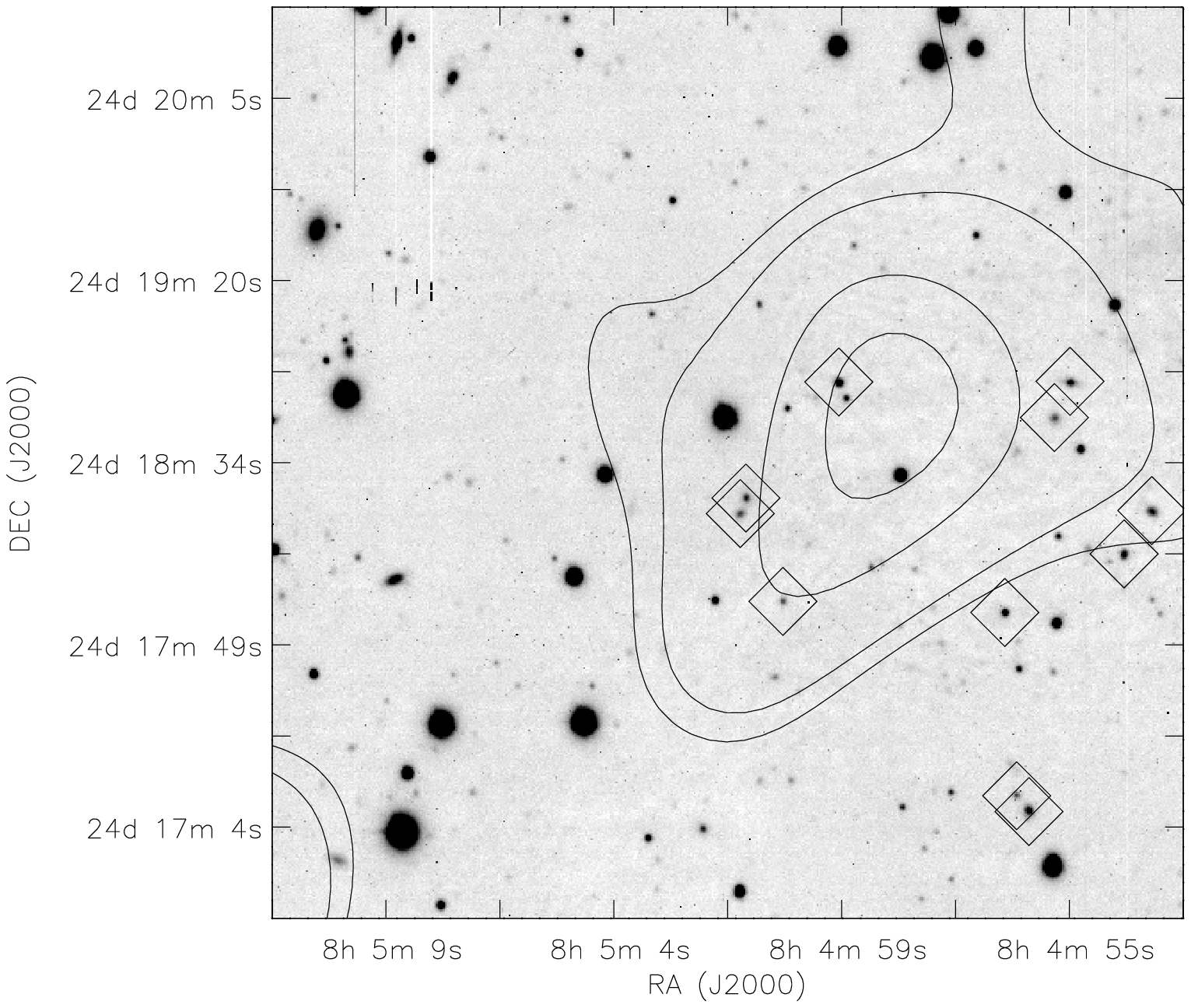}}\\
~~~~~~~~~~~~~~~~{\includegraphics[angle=0, width=0.7\textwidth]{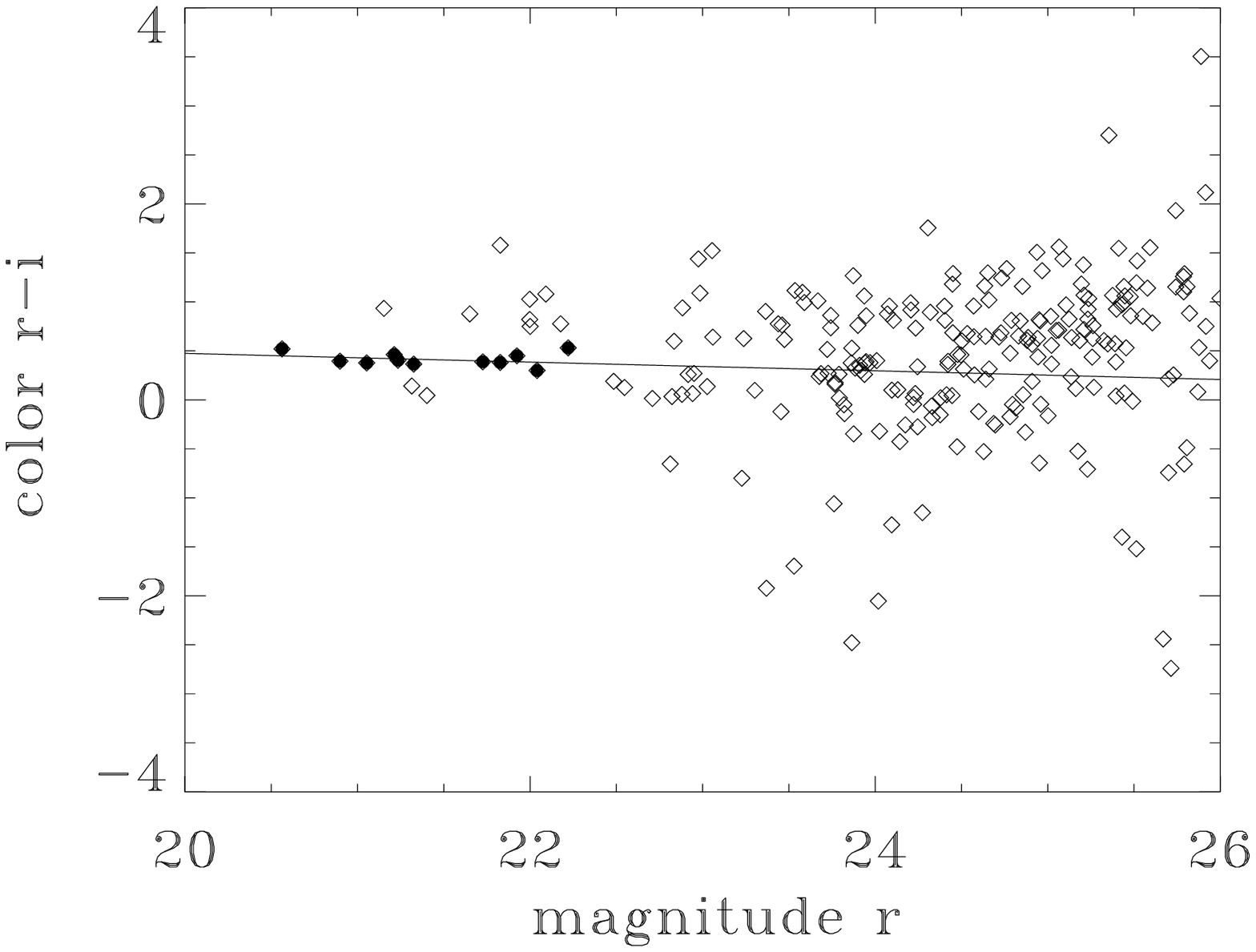}}\\
\end{tabular}
\caption[]{{\bf bot panel:} Colour-magnitude diagram of the cluster 
BMW080459.3$+$241, showing a concentration of the 
brighter galaxies
(solid bullets, $r\,<\,22.5$) around a similar colour. The sky
distribution of these galaxies is marked in the {\bf top panel} over a
100 min image taken with the TNG telescope. The red--sequence colors suggest
a tentative redshift $z\,\simeq\,0.6$.}
\label{eps6}
\end{figure}

We have extracted a sample of cluster candidates based purely on the
source extension (see Fig.~\ref{eps3}), which was requested to exceed the local
PSF at more than $\sim 5\sigma$ level.
These candidates (limited to $\mid b_{II} \mid\,>\,20^\circ$ and with 
flux brighter than $f_x \sim\,3 \cdot 10^{-14}$ erg s$^{-1}$ cm$^{-2}$) have 
been screened by coupling the X-ray isophotes to the
DSS2, discarding 20\%  
of them as obvious contaminants.  
About 90 objects in the current \lq \lq clean'' sample of candidates 
are already identified on the DSS with clear groups or clusters, 
while the remaining $\sim 300$ \lq \lq blank-field'' objects need dedicated 
imaging follow-up work.
This is currently underway using the TNG and ESO 3.6~m telescopes and we
show one example in Fig.~\ref{eps6}.  It is among these \lq \lq invisible'' 
objects that the highest-redshift clusters in this sample are certainly
hidden.  
The success rate of the identification program is so far
high, with 80\% 
of the $\sim 20$ targets observed providing a positive identification. 

\begin{figure}[!t]
\centering
\includegraphics[angle=0, width=.7\textwidth]{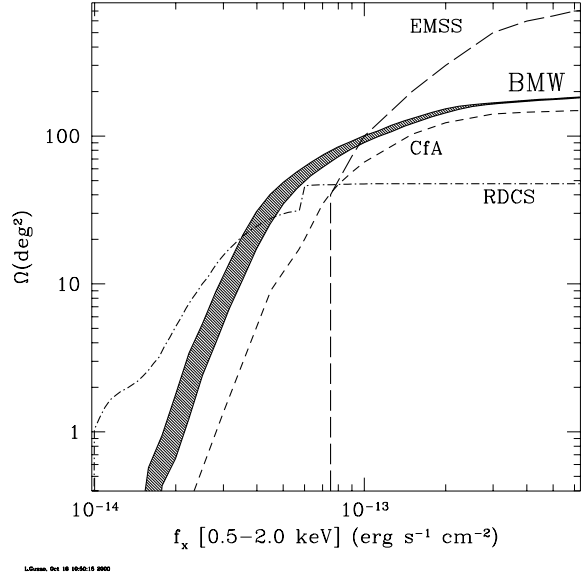}
\caption[]{The solid angle covered by the BMW cluster
sample at different fluxes 
compared to the the EMSS survey (Gioia et al. 1990) and two PSPC surveys 
(Cfa: Vikhlinin et al. 1998b; RDCS: Rosati et al. 1998).}
\label{eps7}
\end{figure}

A quantitative comparison of the BMW cluster sample
properties to those of existing surveys is given in Fig.~\ref{eps7}, 
where the sky coverage at different flux limits is plotted.  
The large sky coverage, nearly 3 times that of the CfA PSPC
sample, makes the BMW cluster sample an excellent
source for finding luminous clusters and thus verify the indications for
evolution of the XLF bright-end yielded by the EMSS and PSPC samples.
\begin{figure}[!t]
\begin{tabular}{cc}
{\includegraphics[angle=270, width=.8\textwidth]{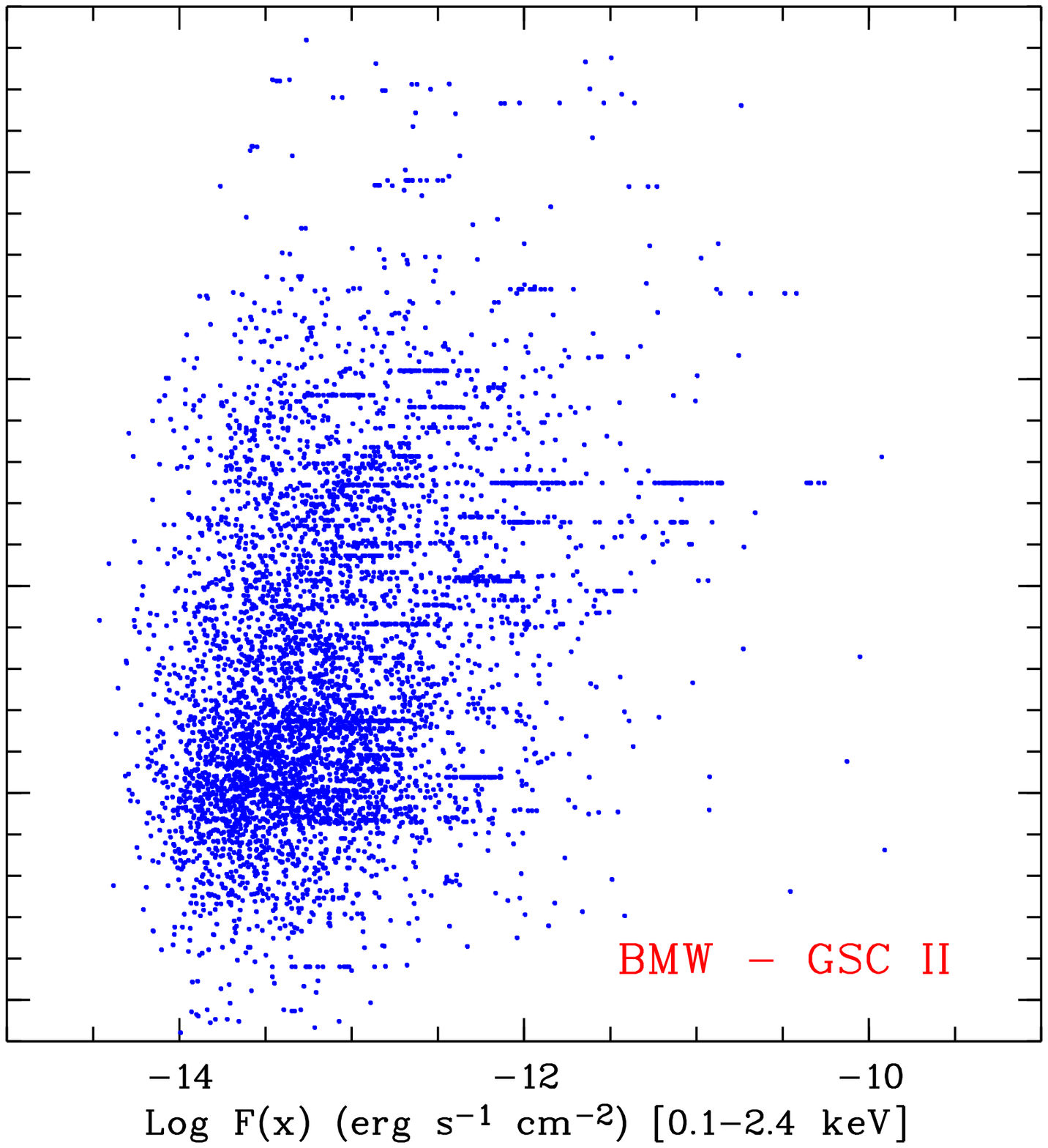}}&{\includegraphics[angle=270, width=.8\textwidth]{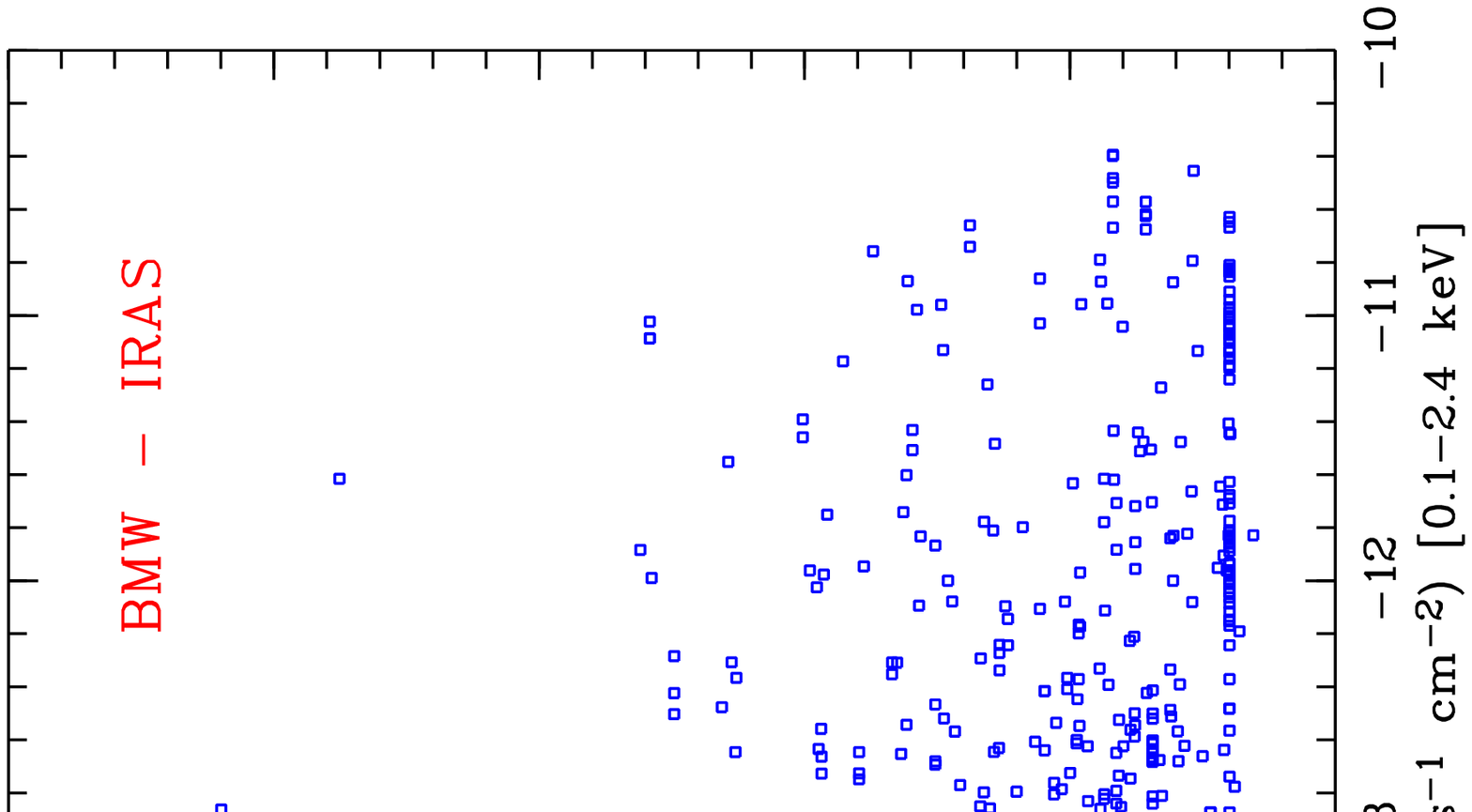}}\\
{\includegraphics[angle=270, width=.8\textwidth]{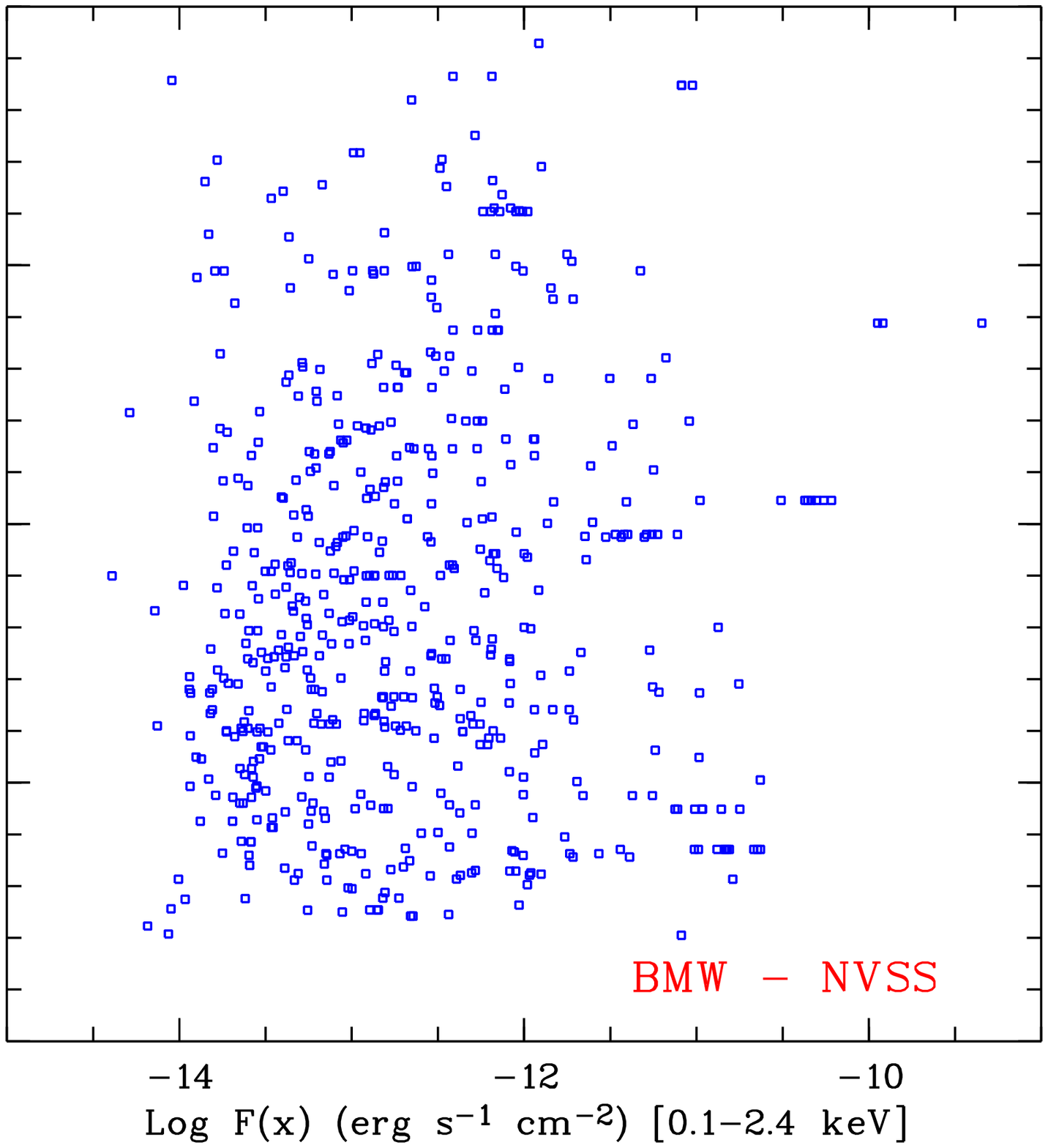}}&{\includegraphics[angle=270, width=.8\textwidth]{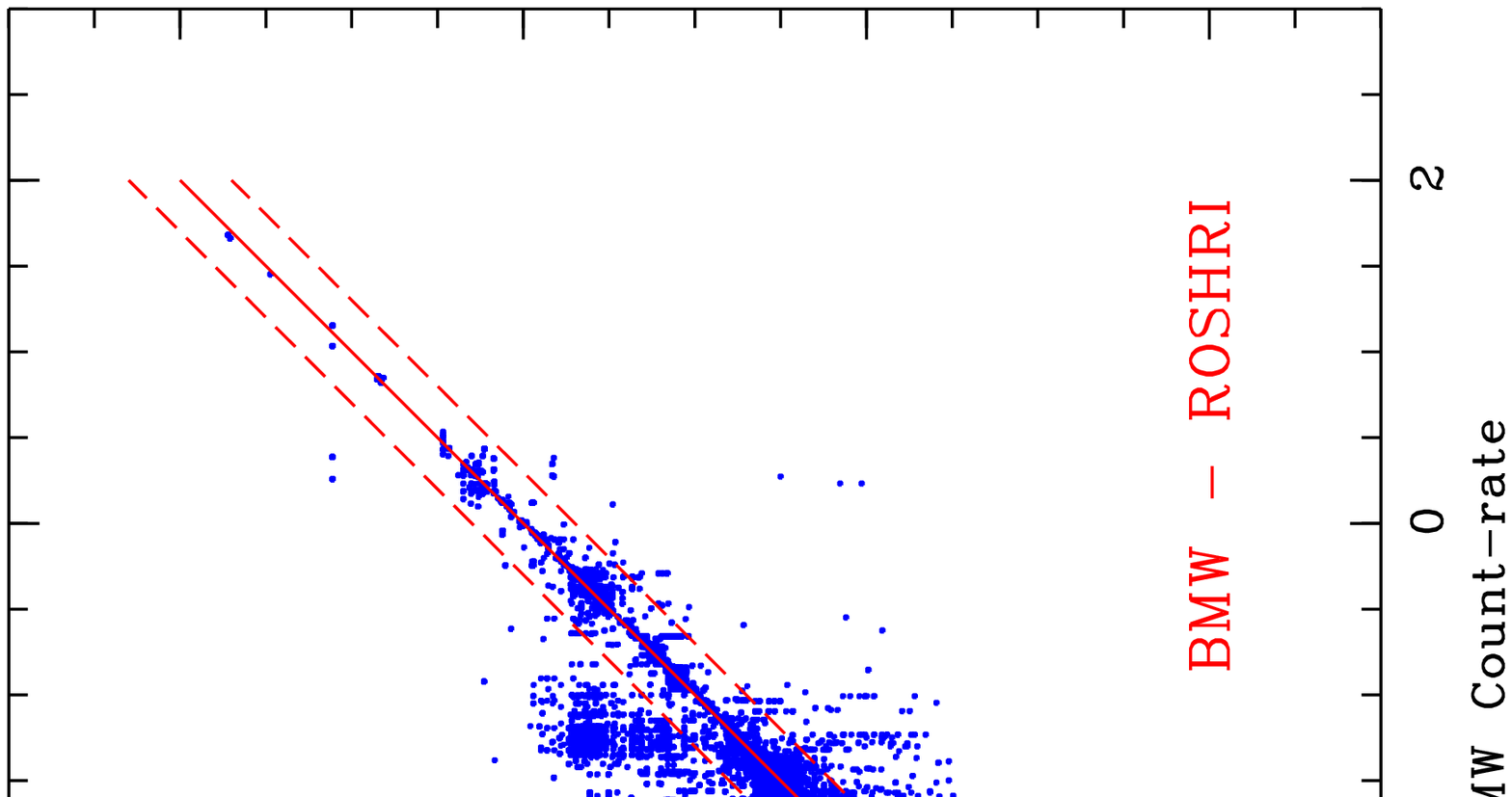}}\\
&\\
\end{tabular}
\caption[]{{\bf upper left panel}: GSC-II B magnitude versus the BMW X-ray 
flux (0.1--2.4 keV) for the $\sim\,10000$ cross-correlated objects 
(10 arcsec cone radius); 
{\bf upper right panel}: IRAS flux (at $12\,\mu$m) versus the BMW X-ray 
flux for the $\sim\,400$ cross-correlated objects 
(1 arcsec cone radius); 
{\bf lower left panel}: NVSS flux (at $20$ cm) versus the BMW X-ray 
flux for the $\sim\,500$ cross--correlated objects (1 arcsec cone radius); 
{\bf lower right panel}: ROSHRICAT count rate versus the BMW count rate for 
the $\sim\,6400$ cross-correlated objects (1 arcsec cone radius).
The two dashed lines represent the region in which the count rate is within a 
factor of two.}
\label{eps11}
\end{figure}
\subsection{Cross-correlations}
The sharp core of the ROSAT HRI Point Spread Function ($< 6$
arcseconds FWHM on-axis) greatly simplifies the search of counterparts 
at different wavelengths, providing valuable information for source 
identification.
As first examples we present in Fig.~\ref{eps11} the cross-correlations of the 
BMW catalogue with the following catalogue: (i)
a preliminary version of the Guide Star Catalog II (GSC--II);
(ii) the Infrared IRAS point source catalogue;
(iii) the NVSS catalogue at the radio wavelenghts;
(iv) the ROSAT source catalogue of HRI pointed observations 
(ROSHRICAT) delivered by the ROSAT Consortium.

The cross-correlation with the GSC-II is still preliminary since the optical
catalogue does not cover all the sky yet. 
As soon as the GSC-II will be completed the cross-correlation will be 
repeated as well.
The cross-correlations of the BMW catalogue with the four catalogue 
provide the following results: $\sim\, 10,000$ cross-correlated objects with 
the GSC-II, $\sim\,400$ with the IRAS catalogue, $\sim\,500$ with the NVSS 
catalogue and finally $\sim\,6400$ with the ROSHRI catalogue.

\subsection{The On-line Service}
A WEB based browser to access the service of the BMW catalogue has been 
developed at the Osservatorio Astronomico di Brera. The source by
coordinate environment allows the search by object name or
coordinates and to choose the output format (table only or table and
sky chart). 
The WEB site can be found at:

\noindent
{\tt http://vela.merate.mi.astro.it/$^\sim$xanadu/browser/sbmw.html}.

\noindent
If you need to get in touch with us write to: {\tt xanadu@merate.mi.astro.it}.

%

\clearpage
\addcontentsline{toc}{section}{Index}
\flushbottom
\printindex


\begin{thebibliography}{7}
%
\addcontentsline{toc}{section}{References}

\bibitem{ref1.1} Campana, S., Lazzati, D., Panzera, M.~R. et al. (1999)
The Brera Multiscale Wavelet ROSAT HRI Source Catalog. II. Application to the
HRI and First Results. ApJ {\bf524}, 423--433

\bibitem{ref1.2} Gioia, I.~M., Maccacaro, T., Schild, R.~E. et al. 1990
The Einstein Observatory Extended Medium-Sensitivity Survey. 
I - X-ray data and analysis. ApJS {\bf72}, 567--619

\bibitem{ref1.3} Israel, G.~L., Panzera, M.~R., Campana, S. et al. (1999)
The Discovery of 321 s Pulsations in the ROSAT HRI Light Curves of 
1BMW J080622.8$+$152732 = RX J0806.3$+$1527.
A\&A {\bf349}, L1--L4

\bibitem{ref1.4} Israel, G.~L., Treves, A., Stella, L. et al. (1998)
First Results from a Systematic Search for New X-ray Pulsators 
in ROSAT PSPC Fields. In: The Many Faces of Neutron Stars, NATO-ASI 
Series, Vol. {\bf515}, Kluwer Academic Publishers, 411--417

\bibitem{ref1.5} Israel, G.~L., Stella, L. (1996)
A New Technique for the Detection of Periodic Signals in \lq \lq Colored'' 
Power Spectra. ApJ {\bf468}, 369--379 

\bibitem{ref1.6} Lazzati, D., Campana, S., Rosati, P. et al. (1999)
The Brera Multiscale Wavelet (BMW) ROSAT HRI Source Catalog. I. The Algorithm. 
ApJ {\bf524}, 414--422

\bibitem{ref1.7} Rosati, P., Della~Ceca, R., Norman, C. et al. (1998)
The ROSAT Deep Cluster Survey: The X-Ray Luminosity Function Out to z=0.8.
ApJ {\bf492}, L21--L24

\bibitem{ref1.8} Rosati, P., Della~Ceca, R., Burg, R. et al. (1995)
A first determination of the surface density of galaxy clusters at very 
low x-ray fluxes. ApJ {\bf445}, L11--L14

\bibitem{ref1.9} Vikhlinin, A., McNamara, B.~R., Forman, W. et al. (1998a)
Evolution of Cluster X-Ray Luminosities and Radii: Results from the 160 
Square Degree ROSAT Survey. ApJ {\bf498}, L21--L25

\bibitem{ref1.10} Vikhlinin, A., McNamara, B.~R., Forman, W. et al. (1998b)
A Catalog of 200 Galaxy Clusters Serendipitously Detected in the ROSAT PSPC 
Pointed Observations. ApJ {\bf502}, 558--581
%
\end{thebibliography}
\end{document}